\documentstyle[prd,eqsecnum,aps,amsfonts]{revtex}

\begin{document}
\draft
\title{Counterexample to the passive topological censorship of
$K(\pi,1)$ prime factors}
\author{Gregory A. Burnett}
\address{Department of Physics, University of Florida,
Gainesville, Florida\ \ 32611}
\date{5 April 1995}
\twocolumn[
\maketitle
\widetext
\begin{abstract}
A globally hyperbolic asymptotically flat spacetime is presented
(having non-negative energy density and pressures) that shows that not
all $K(\pi,1)$ prime factors of the Cauchy surface topology are
passively censored according to asymptotic observers in contradiction
to an argument of Friedman, Schleich, and Witt.
\end{abstract}
]
\narrowtext

\section{Introduction}

One of the intriguing features of Einstein's theory of gravity is that
it allows for the possibility that our Universe is topologically
non-Euclidean.  For instance, somewhere a wormhole may link otherwise
very distant regions of space together.  Why then have such structures
not been observed?  One possibility is that our Universe just happens
to be topologically trivial.  However, Friedman, Schleich, and Witt
argue that even if the Universe is topologically nontrivial, observers
that remain in the asymptotic region (i.e., do not venture into black
holes) cannot probe this topology \cite{FSW93,Friedman91}.  They name
this effect {\it topological censorship}.

Before discussing their results, we first present a few definitions.
Fix an asymptotically flat spacetime $(M,g_{ab})$ and denote the null
infinity associated with an asymptotic region of this spacetime by
${\cal J}$ (which is the disjoint union of a past part ${\cal J}^-$
and a future part ${\cal J}^+$) \cite{Wald84}.  Denote the universal
covering spacetime associated with $(M,g_{ab})$ by
$(\tilde{M},\tilde{g}_{ab})$.  If $M$ is not simply connected, then
$(\tilde{M},\tilde{g}_{ab})$ will have many distinct null infinities
${\cal J}_i$ associated with ${\cal J}$.

An asymptotically flat spacetime $(M,g_{ab})$ is said to satisfy {\it
active topological censorship} if there are no homotopically distinct
causal curves from ${\cal J}^-$ to ${\cal J}^+$ \cite{FSW93}.  In such
a spacetime, there is no way an observer in the asymptotic region can
actively probe the the nontrivial topology of the spacetime by sending
and receiving causal signals as the paths taken by any two such
signals are necessarily homotopic.  If $M$ is simply connected, then
active topological censorship is automatically satisfied.  Otherwise,
active topological censorship is equivalent to the condition that
\begin{equation} \label{active}
J^+({\cal J}^-_i) \cap J^-({\cal J}^+_j) =
\emptyset, \text{ for } i \neq j.
\end{equation}

An asymptotically flat spacetime $(M,g_{ab})$ is said to satisfy {\it
passive topological censorship} if there are no points $p \in M$ for
which there exist homotopically distinct causal curves from $p$ to
${\cal J}^+$ \cite{FSW93}.  In such a spacetime, there is no way an
observer in the asymptotic region can passively observe the the
nontrivial topology of the spacetime by receiving causal signals from
some point in the spacetime as the paths taken by any two such signals
are necessarily homotopic.  Again, if $M$ is simply connected this is
automatically satisfied.  Otherwise, passive topological censorship is
equivalent to the condition that
\begin{equation} \label{passive}
J^-({\cal J}^+_i) \cap J^-({\cal J}^+_j) =
\emptyset, \text{ for } i \neq j.
\end{equation}

Friedman, Schleich, and Witt have proven that active topological
censorship holds for the globally hyperbolic asymptotically flat
spacetimes satisfying the null energy condition \cite{FSW93,ec}.
Therefore, unless our Universe contains matter violating this energy
condition or one is willing to venture into a black hole, any
experiment designed to actively detect the Universe's nontrivial
topology by sending and receiving causal signals will fail (in the
sense that all such signals will be homotopic).

They have also shown that passive topological censorship cannot be
expected to hold for all ``reasonable'' spacetimes by giving an
example of a globally hyperbolic asymptotically flat vacuum spacetime
having the spatial topology of ${\Bbb R}P^3$ minus a point that
violates this condition.  Further, combining a theorem due to Schoen
and Yau \cite{SchoenYau81} and the results of Gromov and Lawson
\cite{GromovLawson83}, they argue that any nontrivial topology of a
Cauchy surface due to a $K(\pi,1)$ prime factor \cite{kpi1} must be
surrounded by an apparent horizon ${\cal A}$, from which they conclude
that any violation of passive topological censorship cannot be due to
the $K(\pi,1)$ prime factors.

Were the apparent horizon ${\cal A}$ surrounding a $K(\pi,1)$ factor
always a {\it future} apparent horizon (the expansion of a family of
future-directed outgoing null geodesics normal to ${\cal A}$ being
zero on ${\cal A}$), then their argument would be correct as such
horizons necessarily lie behind event
horizons\cite{Wald84,HawkingEllis73}.  However, the Schoen-Yau theorem
guarantees only that ${\cal A}$ will be {\it either} a future or past
apparent horizon, which does not allow one to conclude that these
factors in the topology are passively censored.

We show that no such argument can be made by constructing a globally
hyperbolic asymptotically flat spacetime satisfying the
dominant-energy condition \cite{ec}, whose Cauchy surfaces are
diffeomorphic to the three-torus less a point ($T^3$ being a
$K(\pi,1)$ manifold), and whose universal covering spacetime violates
Eq.~(\ref{passive}).  While we do not show that all $K(\pi,1)$ prime
factors need not be passively censored, there seems to be no reason
why they should.  Therefore, while ``reasonable'' spacetimes must
satisfy active topological censorship, a non-simply connected
spacetime should not be expected to satisfy passive topological
censorship.

Our conventions are those of Ref.~\cite{Wald84}.  In particular, our
spacetime metrics are such that timelike vectors have negative norms.

\section{The counterexample}
Our counterexample is constructed from a dust-filled $k=0$
Robertson-Walker spacetime and a Schwarzschild spacetime as follows
\cite{Brill83}.  (Similar counterexamples whose Cauchy surface
topology is a closed hyperbolic space less a point can be constructed
{}from the $k=-1$ Robertson-Walker spacetimes.)

The $k=0$ Robertson-Walker spacetimes have ${\Bbb R}^3$ Cauchy
surfaces and can be coordinated so that the metric $h_{ab}$ is given by
\begin{mathletters}
\begin{eqnarray}
h & = & a^2(\eta) (-d\eta^2 + dx^2 + dy^2 + dz^2) \label{cartesian} \\
  & = & a^2(\eta) (-d\eta^2 + d\chi^2 + \chi^2 \Omega),
  \label{spherical}
\end{eqnarray}
\end{mathletters}
where $x$, $y$, and $z$ can be thought of as rectangular coordinates,
$\chi$ can be thought of as a radial coordinate, and $\Omega_{ab}$ is
a unit-metric on the two-sphere.  With dust as a source, $a(\eta) = C
(\eta/2)^2$, for some constant $C$, and the dust flow is orthogonal to
surfaces of constant $\eta$.  We shall take $\eta > 0$.  (Note that
these spacetimes are forever expanding to the future.)

{}From such a $k=0$ Robertson-Walker spacetime $({\Bbb R}^4,h_{ab})$, we
construct a spacetime with $T^3$ Cauchy surfaces $({\Bbb R} \times
T^3, h'_{ab})$ by identifying $({\Bbb R}^4,h_{ab})$ under a discrete
group of isometries which, for definiteness, we shall take to be the
three translations along $x$, $y$, and $z$ by a unit amount.  (This
way our spacetime is a ``unit square cell''.)\ \ This spacetime is
locally spherically symmetric so it can be locally coordinated by
$\eta$, $\chi$, and two angular coordinates as in
Eq.~(\ref{spherical}).

\begin{figure}
\input{psfig}
\centerline{\psfig{figure=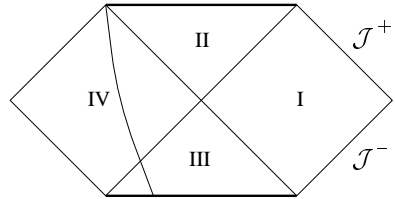}}
\medskip
\caption{A spacetime diagram representing a maximally extended
Schwarzschild spacetime.  The curve extending from region III into
region IV represents the timelike three-surface ${\cal T}'$.  The
surface ${\cal T}'$ is to be attached to its analogous surface ${\cal
T}$ in the dust-filled spacetime with $T^3$ Cauchy surfaces.  The
final spacetime includes everything ``outside'' of ${\cal T}$ in the
$T^3$ Robertson-Walker spacetime and everything ``to the right'' of
${\cal T}'$ in the Schwarzschild spacetime as drawn here.}
\label{schwarzschild}
\end{figure}

We now construct our spacetime $(M,g_{ab})$ as follows.  In the
spacetime $({\Bbb R} \times T^3, h'_{ab})$, consider the (locally
spherically symmetric) timelike three-surface ${\cal T}$ given by
$\chi = \chi_0$ where $\chi_0$ is a constant satisfying $0 < \chi_0 <
1/2$ that will be chosen later.  (The upper bound is needed so that
this tube ``lies within a single cell''.)\ \  We now take the region
lying {\em outside} of ${\cal T}$ and attach it to a Schwarzschild
spacetime of mass $(C/2) \chi_0^3$ by constructing a ``similar''
timelike three surface ${\cal T}'$ therein.  See
Fig.~\ref{schwarzschild}.  (The choice of the mass of the
Schwarzschild spacetime is found by viewing the Robertson-Walker
spacetime as a spherically symmetric spacetime, constructing the mass
function $m$ thereon \cite{Burnett}, and then evaluating $m$ on ${\cal
T}$.  Further, we claim that such a matching can always be done in
such a way that there are no distributional surface layers of matter,
i.e., the matching is $C^1$.)

Our spacetime $(M,g_{ab})$ is globally hyperbolic having Cauchy
surfaces diffeomorphic to $T^3$ minus a point.  Further, the
asymptotic region of this spacetime is the asymptotic region I of the
Schwarzschild spacetime in Fig.~\ref{schwarzschild}.  Although this
spacetime is not vacuum everywhere, it does satisfy the
dominant-energy condition, and indeed just about any reasonable energy
condition as the dust has positive energy density and zero pressure.
We now show that for an appropriate choice of $\chi_0$, the spacetime
$(M,g)$ does not satisfy passive topological censorship.  To do this,
we construct its universal covering spacetime and then construct the
sets $J^-({\cal J}^+_i)$.

\begin{figure}
\input{psfig}
\centerline{\psfig{figure=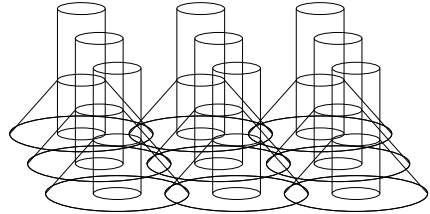}}
\medskip
\caption{A 2+1 dimensional version of a portion of the universal
covering spacetime of $(M,g_{ab})$.  Each of the (locally spherically
symmetric) timelike three-surfaces are attached to identical (though
distinct) Schwarzschild spacetimes.  A point $p$ that lies in the
overlap region of two downward ``cones'' can reach distinct null
infinities by a future-directed causal curve.  Therefore, there exist
homotopically distinct future-directed causal curves from $p$ to
${\cal J}^+$.}
\label{universal}
\end{figure}

The universal covering spacetime $(\tilde{M},\tilde{g}_{ab})$ of
$(M,g_{ab})$ can be represented by a $k=0$ Robertson-Walker spacetime
with a Schwarzschild spacetime attached to each timelike three-surface
of ``radius'' $\chi_0$ with ``center'' having integer $x$, $y$, $z$
coordinates.  See Fig.~\ref{universal}.

\begin{figure}
\input{psfig}
\centerline{\psfig{figure=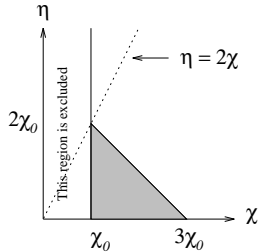}}
\medskip
\caption{A spacetime diagram showing the portion of $J^-({\cal
J}^+_i)$ in the Robertson-Walker part of the spacetime (i.e., the
region shaded).  The vertical line at $\chi=\chi_0$ represents ${\cal
T}_i$.  The region of interest is precisely the region that is both
``outside'' of ${\cal T}_i$ and that lies to the past of the portion
of ${\cal T}_i$ on which $\nabla^a r$ is past-directed timelike.  (In
this diagram, each point represents a sphere, excepting along the line
$\chi = 0$.)}
\label{Robertson-Walker}
\end{figure}

Consider the null infinity ${\cal J}_i$ associated with one of the
asymptotic Schwarzschild regions, e.g., the one ``centered'' at
$(x,y,z) = (0,0,0)$.  From Fig.~\ref{schwarzschild}, we see that the
portion of $J^-({\cal J}^+_i)$ in the Robertson-Walker part of the
spacetime is equivalent to the past of the portion of ${\cal T}'_i$
which lies in region~III of the Schwarzschild spacetime in
Fig.~\ref{schwarzschild}.  This is the portion of ${\cal T}'_i$
(equivalently ${\cal T}_i$) where $\nabla^a r$ is past-directed
timelike, where $r$ is the size of the spheres of symmetry, which in
our case is given by $r = a(\eta) \chi$.  A simple calculation shows
that $\nabla^a r$ is past-directed timelike on the region where $\eta
< 2\chi$.  Therefore, the region were are interested in is that shaded
in Fig.~\ref{Robertson-Walker}.  For a point $p$ in this region, there
exists a future-directed causal curve passing through ${\cal T}_i$
that connects $p$ to ${\cal J}^+_i$.

So, since our ``Schwarzschild ends'' in the universal covering
spacetime are ``separated by a unit amount'', from
Fig.~\ref{Robertson-Walker} it is clear that if we construct our
spacetime with $\chi_0 > 1/6$, the ``shaded regions'' will overlap,
and therefore, as illustrated in Fig.~\ref{universal}, the pasts of
distinct future null infinities will overlap.  Therefore, condition
Eq.~(\ref{passive}) is violated for $\chi_0 > 1/6$, showing that these
spacetimes violate passive topological censorship.

\acknowledgements
I would like to thank John Friedman and Donald Witt
for their comments and for answering my many questions.


\begin{references}
\bibitem{FSW93} J. L. Friedman, K. Schleich, and D. M. Witt, Phys.\
Rev.\ Lett.\ {\bf 71}, 1486 (1993).
\bibitem{Friedman91} J. L. Friedman, in {\it Conceptual Problems of
Quantum Gravity}, edited by A. Ashtekar and J. Stachel (Birkhauser,
Boston, 1991).
\bibitem{ec} A spacetime with Einstein tensor $G_{ab}$ is said to
satisfy the null energy condition (also known as the null-convergence
condition) if $G_{ab}k^ak^b \ge 0$ for all null $k^a$ Actually, they
prove their theorem under a slightly weaker condition known as the
averaged null energy condition.  The dominant-energy condition is said
to be satisfied if $G_{ab}t^au^b \ge 0$ for all future-directed
timelike $t^a$ and $u^b$.
\bibitem{Wald84} R. M. Wald, {\it General Relativity} (University of
Chicago Press, Chicago, 1984).
\bibitem{SchoenYau81} R. Schoen and S. -T. Yau, Commun.\ Math.\ Phys.\
{\bf 79}, 231 (1981).
\bibitem{GromovLawson83}
M. Gromov and H. B. Lawson, Jr., Inst.\ Hautes Etudes Sci.\
Publ.\ Math.\ {\bf 58}, 83 (1983).
\bibitem{kpi1} For a discussion of the prime factorizations of
three-manifolds, see J. Hempel, {\it 3-Manifolds} (Princeton
University Press, Princeton, 1976).  A closed manifold $\Sigma$ is a
$K(\pi,1)$ manifold, for some group $\pi$, if $\pi_1(\Sigma) = \pi$
(i.e., the fundamental group of $\Sigma$ is the group $\pi$) and
$\pi_k(\Sigma) = 0$ for $k \neq 1$ (i.e., all homotopy groups of
$\Sigma$ higher than the first are trivial).  Equivalently, a closed
manifold $\Sigma$ is a $K(\pi,1)$ manifold if its universal covering
manifold is deformable to a point.  For example, the three-torus $T^3$,
having ${\Bbb R}^3$ as its universal cover, is a $K(\pi,1)$ manifold.
\bibitem{HawkingEllis73} S. W. Hawking and G. F. R. Ellis, {\it The
Large-Scale Structure of Space-time} (Cambridge University Press,
Cambridge, England, 1973).
\bibitem{Brill83} D. R. Brill, in {\it Proceedings of the Third Marcel
Grossmann Meeting on the Recent Developments of General Relativity},
edited by H. Ning (North Holland, New York, 1983).
\bibitem{Burnett}
G. A. Burnett, Phys.\ Rev.\ D {\bf 48}, 5688 (1993).
\end{references}
\end{document}